%% file: main.tex
\documentclass[sigconf]{acmart}

\AtBeginDocument{%
  }

\newcommand {\pipelinename}{RBM}

\copyrightyear{2026}
\setcopyright{cc}
\setcctype{by}
\acmConference[ACM UIST '26]{The 39th Annual ACM Symposium on User Interface Software and Technology}{November 02--05, 2026}{Detroit, MI, USA}
\acmBooktitle{The 39th Annual ACM Symposium on User Interface Software and Technology (UIST '26), November 02--05, 2026, Detroit, MI, USA}
\acmDOI{10.1145/3830398.3830657}

\usepackage{subcaption}

\usepackage{url}
\usepackage{xspace}
\usepackage{placeins}

\begin{document}

\makeatletter
\def\@mktitle@iii{%
  \hsize=\textwidth
  \setbox\mktitle@bx=\vbox{%
    \centering

    \parbox{0.90\textwidth}{%
      \centering
      \normalfont\fontsize{9.5pt}{11pt}\selectfont\itshape
      This is the author's version of the article
      that has been accepted by ACM UIST 2026\\[-0.05em]
      and will appear in the Proceedings of the 39th Annual ACM Symposium
      on User Interface Software and Technology.
    }\par

    \vspace{1.15em}

    \@titlefont\centering
    \@ACM@title@width=\hsize
    \parbox[t]{\@ACM@title@width}{%
      \centering
      \@titlefont
      \@title
      \@translatedtitle
      \ifx\@subtitle\@empty\else
        \par\noindent
        {\@subtitlefont
        \@subtitle
        \@translatedsubtitle}%
      \fi
    }%
    \par\bigskip
  }%
}
\makeatother
\title{Reference-Based Manipulation: A Design Space and System for Speech-Driven Spatial Reasoning in VR}

\title{Reference-Based Manipulation: A Framework and Pipeline for Multimodal Spatial Reasoning}


\author{Yangyang He}
\authornote{Both authors contributed equally to this research.}
\orcid{0009-0009-1246-3110}
\affiliation{%
  \department{School of Industrial Design}
  \institution{Georgia Institute of Technology}
  \city{Atlanta}
  \state{GA}
  \country{USA}}
\email{yangyanghe0821@gmail.com}

\author{Zhuangze Hou}
\orcid{0009-0001-6387-2506}
\authornotemark[1]
\affiliation{%
  \department{School of Creative Media}
  \institution{City University of Hong Kong}
  \city{Hong Kong}
  \country{China}}
\email{zhuanghou3-c@my.cityu.edu.hk}

\author{Yonglin Chen}
\orcid{0009-0004-3081-1813}
\affiliation{%
  \department{School of Design}
  \institution{Southern University of Science and Technology}
  \city{Shenzhen}
  \country{China}}
\email{yonglin0711@gmail.com}

\author{Can Liu}
\orcid{0000-0003-3267-3317}
\authornote{Corresponding author.}
\affiliation{%
  \department{School of Creative Media}
  \institution{City University of Hong Kong}
  \city{Hong Kong}
  \country{China}}
\email{canliu@cityu.edu.hk}

\renewcommand{\shortauthors}{He et al.}

\begin{abstract}
When manipulating objects in immersive platforms through speech and gesture, users naturally construct spatial references—referring to scene entities, their bodies, or the environment. Leveraging spatial cognition theories, this work systematically examines how users construct and communicate spatial intent.
Using a custom toolkit, we conducted a Wizard-of-Oz study to observe unconstrained multimodal (speech + gesture) input patterns in Virtual Reality for scene construction. Based on these findings, we formalize a framework that decomposes spatial references into three core components—Source, Anchor, and Frame—while characterizing their compositional strategies and explicitness. We demonstrate the utility of this Reference-based Manipulation framework by implementing an LLM-based pipeline featuring a set of example interaction techniques with a preliminary technical evaluation. Finally, we discuss key lessons learned for supporting reference-based spatial interaction.

\end{abstract}





\begin{teaserfigure}
    \centering
    \includegraphics[width=1\textwidth]{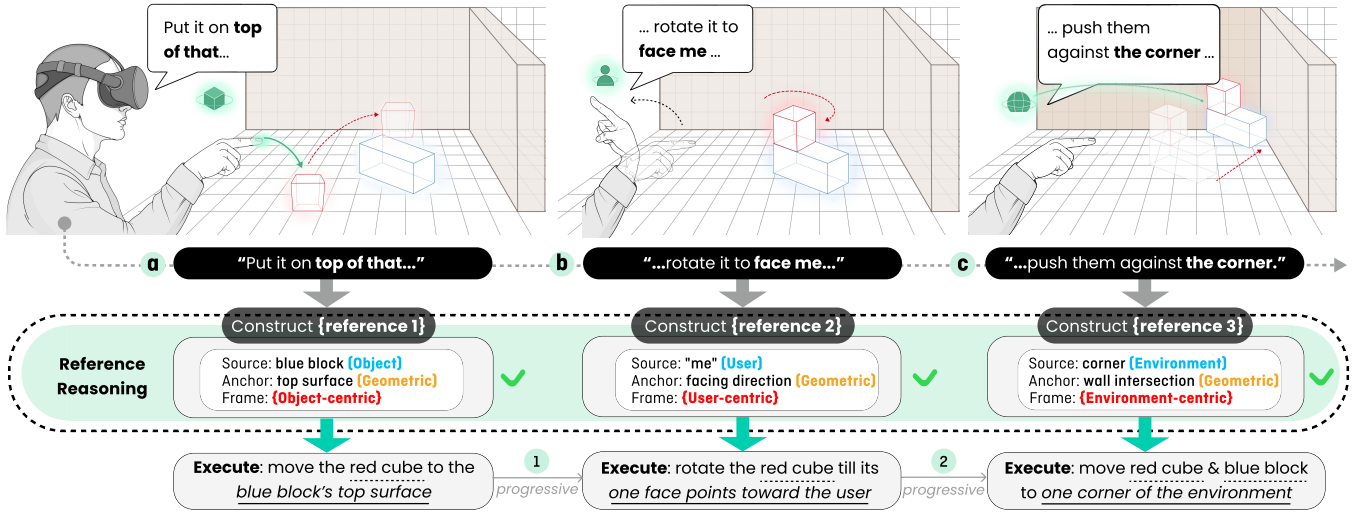}
    \caption{
    User uses natural speech and gesture to manipulate objects in a VR scene. Under Reference Reasoning (green), \pipelinename{} pipeline decomposes each instruction into $r$(\textit{Source}, \textit{Anchor}, \textit{Frame}): (a)~gesture pointing resolves the implicit Source ``that'' to the blue block; (b)~the user's body orientation provides the Anchor for ``face me''; (c)~first-person scene analysis identifies the environment corner. The three turns form a progressive multi-reference chain, with each new reference building on the last.}
    \Description{A three-part teaser figure. The top row shows three VR scene illustrations of a user issuing progressive speech-and-gesture commands: (a) pointing at a blue block while saying ``Put it on top of that,'' (b) gesturing toward themselves while saying ``rotate it to face me,'' and (c) selecting two objects while saying ``push them against the corner.'' Below the illustrations, a gray Command Paradigm row shows three failed function calls with unresolved parameters marked by red crosses. A green Reference Reasoning row shows RBM decomposing each instruction into Source, Anchor, and Frame, with green checkmarks. At the bottom, an Execute row displays the resolved operations connected by dashed arrows labeled ``progressive.''}
    \label{fig:teaser}
\end{teaserfigure}

\maketitle

\input{1-Intro}
\input{2-RelatedWorks}

\input{3-Theoretical_Foundations}
\input{4-UserStudy}

\input{5-Design_Sapce}

\input{6-Pipeline}
\input{9-Conclusion}

\bibliographystyle{ACM-Reference-Format}
\bibliography{references}
\input{10-Appendix}





\end{document}

%% file: 1-Intro.tex
\section{Introduction}
Manipulating 3D objects in virtual reality (VR) demands that users communicate complex spatial intent: where to place an object, how to orient it, how large to make it, and \textit{relative to what}. Traditional manipulation interfaces require users to perform sequences of selection, transformation, and confirmation actions using controllers, gizmos, or direct hand interaction~\cite{mendes2019survey, laviola20173d, yu2021gaze, hand1997survey, bowman2001introduction}. While precise, these interfaces impose a learning curve and can induce arm fatigue during prolonged use~\cite{jang2017modeling, hincapie2014consumed}. More critically, they require users to translate their spatial intentions into a vocabulary of coordinate axes, rotation handles, and numerical parameters that bears little resemblance to how humans naturally think and talk about space.

In everyday life, people communicate spatial manipulation instructions effortlessly by combining speech and gesture. We say ``put it \underline{next to} the table'' while pointing, or ``make it the same size as \underline{that one},'' relying on the listener's understanding of what we are referring to. This capacity for multimodal spatial communication has long been recognized in HCI, from Bolt's seminal ``Put-that-there''~\cite{bolt1980put} to recent explorations of eye-hand coordination~\cite{wagner2024eye}, gaze-assisted selection~\cite{yu2021gaze, pfeuffer2014gaze}, hand interfaces \cite{pei2022hand}, and co-speech gesture~\cite{williams2020understanding} in immersive environments. Motivated by these findings, a new generation of LLM-powered systems has begun to support speech-based object manipulation in VR and AR~\cite{wang2025can, hu2025gesprompt, zhang2024vrcopilot, hou2025echoladder, chen2025analyzing, zhu2025agentar}. These systems allow users to issue spoken commands such as ``\underline{move} the chair \underline{to} the window'' and use an LLM to parse the utterance into executable operations.

Established theories of spatial cognition~\cite{levinson2003space, klatzky1998allocentric} and spatial language~\cite{tversky2005functional, talmy2000toward} suggest that human expressions of spatial intent are inherently referential. 
Yet, to our knowledge, no prior work systematically analyzed and integrated \textit{how the user constructs and communicates spatial intent} with multimodal input. 
We argue that the construct underlying spatial commands is not the \textit{command} but the \textit{reference}: a relational, anchored spatial constraint that users construct by invoking entities in the scene, extracting properties from them, and interpreting the result through a cognitive coordinate system. 
Consider three successive instructions a user might give while arranging objects in an immersive virtual space: ``put it \underline{on top of} that,'' ``rotate it to \underline{face me},'' and ``push them against \underline{the corner}'' (Figure~\ref{fig:teaser}). This single sequence invokes three different reference sources (a scene object, the user's body, the environment), extracts different properties from each (a surface, an orientation, a contact position), interprets each through a different cognitive coordinate system (the object's intrinsic axes, the user's body, the scene's stable structure), and composes them progressively across turns. 
With the increased use of natural language processing in new operating systems, we believe that supporting such operations in VR/MR could open up new possibilities for embodied interfaces.

To empirically understand how users express spatial intent, we conducted a Wizard-of-Oz study (N=12) in which participants performed unconstrained speech-and-gesture manipulation across three VR scene building tasks. 
The study yielded 903 executed user instructions, corresponding to 1,436 system operations logged by the WIZVR toolkit. Analysis of these operations
shows that referential expression is pervasive, 
with users invoking scene objects, their own bodies, the environment, and even prior states as referenced entities. 

Building on these theoretical and empirical foundations, we introduce \textit{Reference-based Manipulation}, comprising: (1)~a \textbf{framework} that decomposes any spatial reference into three components, \textit{Source} (entity providing reference), \textit{Anchor} (property extracted), and \textit{Frame} (cognitive coordinate system), and characterizes how multiple references are organized along two orthogonal dimensions (\textit{Strategy} and \textit{Explicitness}); and (2)~a \textbf{reference-reasoning pipeline} supporting a set of demo interaction techniques 
using chain-of-thought LLM inference to identify, ground, and execute spatial references from multimodal input.

In summary, the contributions of this work are threefold:

\begin{enumerate}
    \item \textbf{Referential patterns} observed from 903 user instructions 
    (involving 1,436 system operations) 
    collected from a Wizard-of-Oz study (N=12) 
    in VR, including how they are organized around object, user, and environment references and how they invoke prior states as temporal anchors.
    \item A \textbf{Reference Interpretation Framework} formalizing a Reference Construction Model $r$(\textit{Source}, \textit{Anchor}, \textit{Frame}) and a design space of compositional strategies across references. 
    \item An \textbf{LLM-based pipeline} instantiating the framework, accompanied by 
    implementation lessons
    and a preliminary technical evaluation 
    of its feasibility, robustness, and current limitations.

\end{enumerate}

%% file: 2-RelatedWorks.tex
\section{Related Works}

We review three bodies of work that collectively motivate our approach. 
Table~\ref{tab:comparison} provides a comparative summary of recent speech-based multimodal interfaces for spatial interaction in XR. 
 To our knowledge, \pipelinename{} is the first system to ground its computational pipeline in a systematic, empirically derived design space of spatial reference, and to support reference reasoning, rather than command classification, as the organizing principle for speech-driven manipulation in VR.

\subsection{Speech and Gesture Integration for 3D Manipulation}
Combining speech and gesture for spatial interaction dates to Bolt's "\textit{Put-that-there}" \cite{bolt1980put}, which demonstrated that deictic pointing and voice jointly produce commands neither modality can express alone. \citet{oviatt1999ten} 
established that these inputs are complementary: spatial content tends toward gesture while relational content tends toward speech---a principle also demonstrated in earlier multimodal architectures such as QuickSet \cite{cohen1997quickset}.
Subsequent elicitation work produced gesture taxonomies and timing profiles for AR manipulation \cite{williams2020understanding, williams2022impacts} following established elicitation methodology \cite{wobbrock2009user}, with Zhou et al. \cite{zhou2022eliciting} finding that speech plays a greater coordinative role with multiple targets.
Participants in Aghel Manesh et al. \cite{aghel2024people} routinely pointed to scene locations, used deictic expressions such as “here” and “that,” and implicitly assumed shared spatial knowledge with the system, revealing pervasive embodied referencing in VR.
Collectively, these works provide rich behavioral descriptions of multimodal input in immersive manipulation contexts.  
However, these studies largely operate at the modality level, describing what users produce without explicitly modeling the underlying structure that organizes these behaviors. In other words, they do not explain why speech and gesture in spatial manipulation take the particular forms they do. As a result, the insights these studies yield remain primarily descriptive and have not been systematically translated into computational principles for system design. A new generation of LLM-based systems has begun to act on multimodal input for scene manipulation, which we review next.

\subsection{LLM-Based Intelligent Scene Manipulation}
Recent advances in LLMs have enabled a new generation of multimodal systems that use natural language to drive VR\&AR scene manipulation. \textit{VRCopilot} \cite{zhang2024vrcopilot} translates speech and pointing into wireframe-based placement;
\textit{GesPrompt} \cite{hu2025gesprompt} extracts spatial-temporal parameters from co-speech gestures to fill LLM function calls; 
\textit{EchoLadder} \cite{hou2025echoladder} generates progressive modification suggestions; 
Chen et al. \cite{chen2025analyzing} analyzed multimodal strategies in LLM-assisted 3D editing. More broadly, recent surveys have mapped the rapid growth of speech-related multimodal interaction in XR \cite{yu2024object}.
Prior systems such as LLMR \cite{de2024llmr} and DreamCodeVR \cite{giunchi2024dreamcodevr} demonstrated LLM-driven scene editing through code generation, though with response times ranging from 15 seconds to several minutes.

While these systems vary in scope, they share a common computational architecture: (1) speech is parsed into intent categories and parameter slots, and (2) gesture, gaze, or scene context is used to fill missing parameters. This command classification and parameter filling paradigm is effective for well-structured instructions, but its pipeline designs are driven primarily by engineering considerations 
rather than by systematic observation of how users naturally express spatial intent through language and gesture. Recent works have begun to address specific aspects of natural language understanding: \textit{GazePointAR} \cite{lee2024gazepointar} resolves ambiguous pronouns through multimodal context, and \textit{VR Mover} \cite{wang2025can} tracks conversational context to support implicit references and follow-up instructions. These demonstrate that attending to natural communication patterns yields improvements, but the underlying architecture remains command-driven. An alternative in which the pipeline is organized around how users construct spatial intent, rather than around command categories, remains unexplored.

\subsection{Spatial Reference and Cognitive Grounding in XR Manipulation}
A smaller but growing body of work has directly engaged with spatial reference or cognitive foundations as a design resource for XR manipulation. Irawati et al. \cite{irawati2005semantic} encoded object relationships as semantic knowledge for VR manipulation.
Kaiser et al. \cite{kaiser2003mutual} demonstrated mutual disambiguation of speech and 3D gesture in AR \& VR, 
showing that cross-modal disambiguation accounted for over 45\% of successful interpretations. This aligns with Clark’s broader account of referring as a collaborative process grounded in shared context \cite{clark1986referring}.
Wang et al. \cite{wang2024push} presented “\textit{Push-That-There},” extending Bolt’s paradigm to tabletop multi-robot manipulation.
These efforts demonstrate that spatial and semantic reference information can meaningfully improve manipulation systems. However, they address isolated aspects of referencing 
without systematically examining how users construct spatial references more broadly.
Beyond the technical contributions discussed above, two of these systems engage with reference and cognition at a deeper level. \textit{GazePointAR} \cite{lee2024gazepointar} is notable in that it treats reference resolution, not just intent classification, as a first-class design problem.
\textit{VR Mover} \cite{wang2025can} grounds its interaction design in visual working memory theory, applying principles of chunking and coarse-to-fine processing to enable effective multi-object manipulation. 
Both demonstrate the value of cognitive grounding but each addresses a single dimension of the referencing problem. 
Yet no existing work provides a unified account of spatial reference construction during multimodal VR interaction. Our work addresses this gap by identifying the components of a reference, characterizing how references are composed and how much is left implicit, and translating these empirical regularities into pipeline architecture. Our work is grounded in established theories of spatial reference in human cognition, which we synthesize in the next section.

%% file: 3-Theoretical_Foundations.tex
\section{Spatial Reference in Human Cognition}
Our work is grounded in theories of spatial cognition, psycholinguistics, and gesture studies. Although not originally formulated for digital platforms, these theories provide fundamental concepts for humans' spatial reasoning. 
We synthesize them into 
three themes. 

\subsection{Frames of Reference in Spatial Cognition}
A central finding of spatial cognition research is that humans do not encode space in a single, unified coordinate system. Instead, people construct spatial meaning through \textit{frames of reference}, which are coordinate systems anchored to different entities in the environment, and dynamically select among them depending on communicative context, task demands, and cultural convention.

\citet{levinson2003space} provides a widely used typology based on cross-linguistic fieldwork. He identifies three frames: the \textit{intrinsic frame}, which defines locations using the reference object's inherent axes (e.g., front, back); the \textit{relative frame}, which defines locations from a viewpoint, typically the speaker's body-centered axes; and the \textit{absolute frame}, which uses fixed environmental axes such as cardinal directions. Critically, Levinson documents \textbf{frame switching} within single communicative episodes: speakers routinely shift frames as they describe different aspects of a scene, and this switching is typically implicit. Subsequent experimental work has confirmed that such frame preferences vary systematically across tasks and spatial configurations \cite{carlson1993frames, mou2002intrinsic}, reinforcing that frame selection is context-dependent rather than fixed.

From a cognitive neuroscience perspective, \citet{klatzky1998allocentric} distinguishes \textit{egocentric} representations (encoding locations relative to the observer's body) from \textit{allocentric} representations (encoding locations relative to external landmarks). These systems operate concurrently and must be coordinated; behavioral evidence indicates a measurable \textbf{switching cost} when shifting between them \cite{klatzky1998allocentric}, reflected in slower responses and higher error rates. For VR interaction, this means misidentifying the user's active frame introduces avoidable cognitive load.
Together, Levinson and Klatzky converge on a shared insight: \textbf{humans maintain multiple, simultaneously available spatial coordinate systems and dynamically select among them}. Systems that assume a single fixed coordinate system will misinterpret a substantial portion of spatial commands.

\subsection{The Relational and Qualitative Nature of Spatial Language}
A second foundational finding concerns the form of spatial language. \citet{tversky2005functional} demonstrated that spatial descriptions are not metric but \textbf{qualitative and relational}: mental representations of space are schematic and topological, preserving functionally relevant structure while abstracting away metric detail. People say ``the cup is next to the keyboard'' rather than specifying centimeters and angles. \citet{tversky1998space} further show that spatial descriptions are perspective-dependent, with speakers shifting between survey and route perspectives as communicative goals change---paralleling Levinson's frame switching. This pattern is robust across different spatial scales and description tasks \cite{taylor1996perspective}.

More broadly, research on spatial semantics has established that spatial language encodes \textit{topological} and \textit{projective} relations rather than metric ones \cite{talmy2000toward}, with the specific interpretation of spatial terms shaped by the geometric and functional affordances of the reference object \cite{landau1993whence}, and that the interpretation of even simple spatial prepositions (``on,'' ``near,'' ``in front of'') depends heavily on the geometric and functional properties of the reference object \cite{herskovits1986language}. The implication for VR interaction design is clear: \textbf{a system should expect qualitative, relational instructions as the default}, inferring metric parameters from qualitative descriptions rather than demanding precise specifications.

\subsection{Gesture and Speech as a Unified Referential System}

The third theoretical pillar concerns how spatial meaning is distributed across modalities. \citet{mcneill1992hand} established that speech and gesture form a single integrated system for expressing thought: they are temporally synchronized, co-expressive, and informationally complementary. Cassell et al. \cite{cassell1999speech} further showed that conversational gestures serve specific discourse functions, including grounding referents in shared space. McNeill identifies \textit{deictic gestures} (grounding referents in shared space) and \textit{iconic gestures} (depicting spatial properties such as shape or orientation) as especially relevant for spatial communication. \citet{kendon2004gesture} provides convergent evidence, showing that gesture is not merely an accompaniment to speech but carries communicative intent in its own right, particularly for spatial and directional information.

\citet{kita2003does} extended this framework by showing that gesture is shaped jointly by spatial thinking and the linguistic structure of accompanying speech---the so-called Interface Hypothesis. The implication is that gesture cannot be treated as a direct readout of spatial cognition; it must be interpreted relative to the concurrent speech. \citet{oviatt1999ten} further supports this with empirical evidence that multimodal input is typically complementary rather than redundant, with spatial content tending toward gesture and relational content toward speech.
This further motivates us to systematically analyze how speech and gesture work together to support spatial referencing and expression, in order to support natural interaction methods for multimodal systems. 

%% file: 4-UserStudy.tex
\section{Wizard-of-Oz Study}
To systematically observe user behaviors in 
spatial reference construction and in immersive space, 
we conducted a Wizard-of-Oz study using a custom VR toolkit, in which participants performed object manipulation tasks using unconstrained speech and gesture. 
Based on the theoretical foundations established in the previous section, we formulate three research questions that guide our empirical investigation. 

\begin{enumerate}
\item[\textbf{(RQ1)}] \textbf{How do users make referential expressions for VR spatial manipulation?}

\item[\textbf{(RQ2)}] \textbf{What strategies do users employ across multiple references?} 
 
\item[\textbf{(RQ3)}] \textbf{How are speech and gestural input used for making referential instructions?}

\end{enumerate}

\begin{figure*}[h]
    \centering
    \includegraphics[width=1.9\columnwidth]{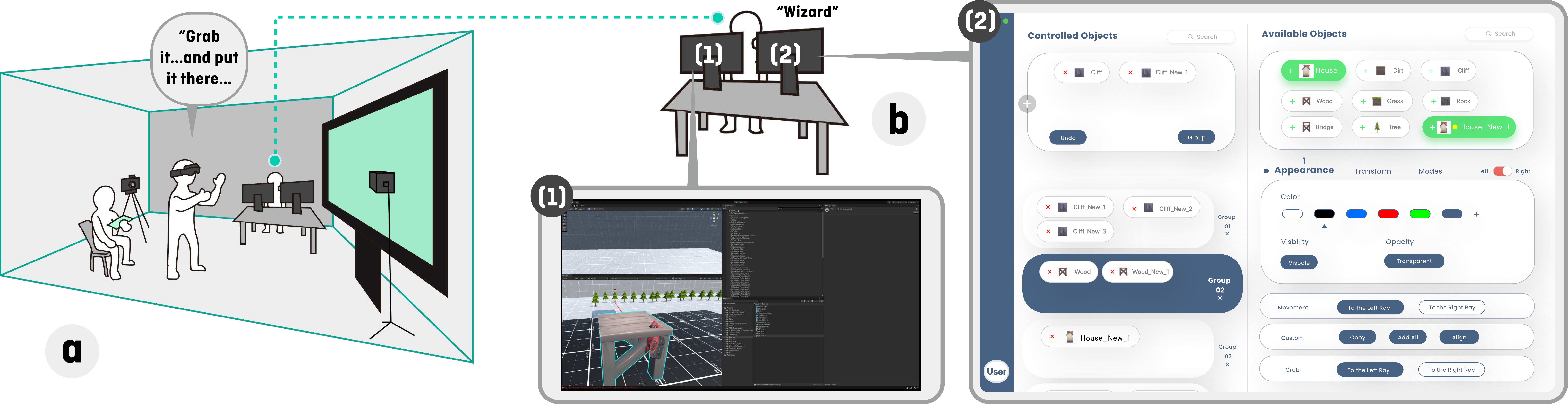}
    \caption{Study setup and WIZVR toolkit. (a)~Participant interacts via speech and gesture; wizard operates WIZVR from a separate station. (b)~Wizard's workspace: monitor~(1) shows the live Unity scene and diagnostics; monitor~(2) mirrors the participant's first-person view; the WIZVR toolkit (right) supports multiple object selection and button control for placement, duplication, alignment and appearance changes, etc.} 
    \Description{A two-panel figure. Panel (a) is an illustrated diagram of the lab setup showing a participant in a VR headset, a wizard seated at a desk with two monitors, a front-facing display, and a camera. Panel (b) shows the wizard's dual-monitor workspace and the full WIZVR toolkit interface with object management panels, color and transform controls, and quick-action buttons.}
    \label{fig:study-setup}
    \vspace{-0.3cm}
\end{figure*}


\subsection{Study Design}
We employed a Wizard-of-Oz (WoZ) approach \cite{dahlback1993wizard,riek2012wizard} to study how users naturally combine speech and gesture for 3D manipulation when no predefined command set or rules are imposed. A hidden human operator (the “wizard”) interpreted participants’ freeform multimodal input and executed the corresponding manipulations in real time, creating the illusion of an intelligent system that could understand any utterance or gesture (Figure~\ref{fig:study-setup}). This method was chosen for two reasons. First, building a fully functional multimodal recognition system would prematurely constrain the interaction vocabulary we aimed to observe. Second, compared with low-fidelity methods such as bodystorming or think-aloud protocols, the WoZ approach enables high-fidelity, near real-time system responses, which are critical for eliciting unconstrained user behavior. 

\subsubsection{Apparatus}
The study was conducted in a controlled lab environment. Participants stood at the center of a tracked space wearing a Meta Quest 3 headset running Unity 3D (Figure~\ref{fig:study-setup}). A large front-facing display mirrored the participant’s first-person VR view for shared observation. The wizard was seated to the participant’s left, operating a custom control toolkit on a laptop connected to two external monitors: one displaying the toolkit interface, the other showing the live VR scene and system diagnostics. An experimenter was positioned behind the participant to provide guidance as needed. Session video and audio were captured using an Insta360 camera and a secondary webcam.

\subsubsection{Custom-made toolkit -- WIZVR}

To support the WoZ approach, we developed the WIZVR toolkit (Fig.~\ref{fig:study-setup}-2) to allow our wizard to perform VR scene manipulation in real time after interpreting the participant's intent. WIZVR presents the wizard with a synchronized view of the participant's first-person perspective across a dedicated monitor for understanding their state, including gaze direction, hand ray intersections, and object highlighting, while a separate display hosts the control interface, separating perception from action.
A central challenge is that the wizard must control a 3D scene through a 2D interface. WIZVR addresses this by exploiting the participant's own embodied input as a spatial bridge: when a participant points at an object, the toolkit highlights the corresponding item in both the VR view and the wizard's panel for unambiguous identification; for deictic placement (e.g., pointing while saying "put it there"), a one-click macro teleports the selected object to the participant's hand-ray coordinates, reducing the wizard's task from 3D spatial reasoning to intent confirmation. The toolkit supports primitive operations on single or multiple objects simultaneously: selection, grab and release, movement, rotation, scaling, color and transparency modification, deletion, and copying. The wizard can also activate gesture-control modes that delegate continuous manipulation (e.g., hand-tracked positioning or two-hand scaling) to the participant, then terminate the mode on command. All actions are relayed as lightweight structured commands through a persistent WebSocket connection to the Unity VR runtime, enabling round-trip latencies sufficient to maintain the illusion of an intelligent system.

\subsubsection{Participants}
We recruited 12 participants (age 19–27, M = 23.75, SD = 2.22; 2 female, 10 male) through university mailing lists and social media. Five had prior VR experience; all reported normal or corrected vision without motor impairment. Participants received a local shopping coupon as compensation valued about \$20 USD. 
The study protocol 
was approved by our Institutional Ethics Review Board (No. HU-STA-00001029).

\subsubsection{Procedure and Tasks}
Each session lasted approximately 100 minutes and comprised three phases. In the \textit{intro phase} (10 min), the experimenter introduced the study and told participants that the system was an intelligent platform capable of interpreting speech and gesture input. Participants were encouraged to use them naturally without relying on conventional VR interfaces. 

In the \textit{task phase} (70 min), participants completed three VR scene construction tasks in fixed order, each lasting approximately 20 minutes.
Participants took short breaks after each task, and the three tasks varied in structure to reduce fatigue.

\textbf{Task 1 (Basic Operations):} Three colored cubes in an empty space. 
This served to familiarize participants with the available operations (move, rotate, scale, color, duplicate, remove) and to establish comfort with freeform speech–gesture manipulation.
\textbf{Task 2 (Structured block-building):} A reference image of a block castle was displayed. Participants were asked to assemble the same structure using the primitive shapes provided in the scene. This ensured a complexity of required manipulation.
\textbf{Task 3 (Open scene creation):} A 
    sandbox with a variety of labeled assets (e.g., houses, trees, and rocks, etc.) for participants to create a scene freely.

In the \textit{post-task phase} (20 min), a semi-structured interview focused on participants’ interaction strategies, their construct of spatial instructions, 
and reflections on the experience.

 \begin{figure*}[h]
   \centering
   \includegraphics[width=\textwidth]{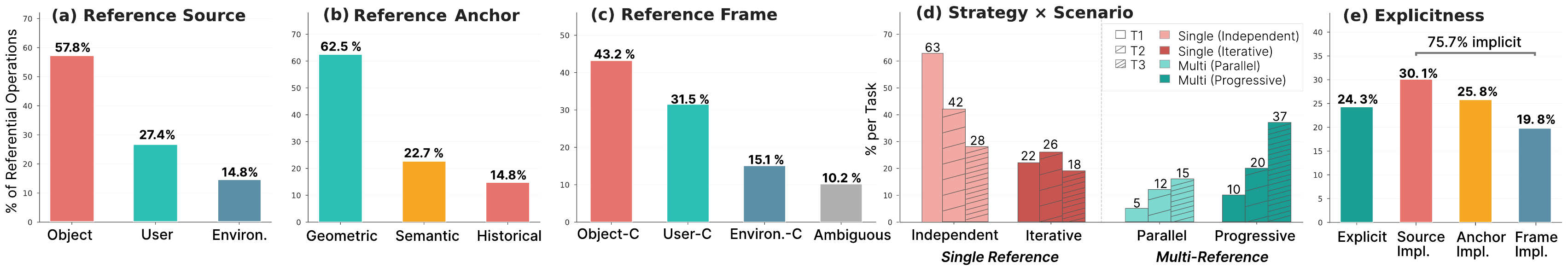}
   \caption{Quantitative distribution of referential behaviors among the referential operations identified within the 1,436 system operations (12 participants).} 
   \label{fig:findings_distribution}
 \end{figure*}

\subsubsection{Data Analysis}
We collected multiple data streams: screen and audio recordings of all sessions, complete WIZVR system logs with timestamped wizard actions, and transcribed participant speech. Our analysis combined quantitative behavioral coding with qualitative thematic analysis \cite{braun2006using}. Three researchers 
first conducted a joint analytic calibration on three task sessions to establish a shared codebook. They then independently coded all sessions, with disagreements resolved through discussion.

Across 12 participants, we identified 903 user instructions,  
corresponding to 1,436 system operations recorded in the WIZVR logs. 
We define a user instruction as a goal-directed unit of user input, expressed through speech, gesture, or their combination, that communicates an intended manipulation and its target(s), either explicitly or implicitly through contextual information. Consecutive micro-adjustments involving the same target(s), the same manipulation type, and the same intent were grouped into a single user instruction, which may involve multiple corresponding system operations. 
For example, successive adjustments such as ``move it left,'' ``a little more,'' and ``slightly back'' were treated as one instruction. 
Repeated inputs while waiting for a system response were also counted as one. 
In contrast, an input that changed the target, manipulation type, or relational constraint was treated as a new user instruction, even when it formed part of a progressive multi-reference sequence. 
We coded each system operation by manipulation type (selection, movement, rotation, scaling, color, duplicating), 
and input modality (speech only, gesture only, speech plus gesture). 
For operations containing referential information, we additionally coded the referenced entity (Source), the extracted property (Anchor), 
the frame of referencing (Frame), and the explicitness of each component.

\subsection{Key Findings}

We analyzed the user instructions and system operations
through the lens of the theoretical framework synthesized in Section~3. 
 Figure~\ref{fig:findings_distribution} summarizes the resulting distributions of referential behaviors.

\subsubsection{\textbf{\textit{Referential expressions constructed with Source, Frame and Anchor (RQ1).}}}

Participant tended to use referential expressions rather than absolute coordinates, consistent with previously mentioned spatial cognition theories ~\cite{levinson2003space, klatzky1998allocentric} 
that humans organize spatial intent through frames of reference.
This pattern held consistently across all three tasks.
Guided by Levinson's tripartite typology of spatial frames~\cite{levinson2003space} and Klatzky's egocentric--allocentric distinction~\cite{klatzky1998allocentric}, we coded each system operation containing referential information for the entity invoked as a \textbf{reference source}, which accounts for over 90\% of the operations. All three source categories appeared across all 12 participants, often within the same task sequence (Figure~\ref{fig:findings_distribution}a).

\paragraph{\textbf{\textit{Reference Sources}}} We identified three types of reference sources.
\textbf{Object as reference}, referencing scene entities, were most frequently observed (57.8\% of the referential operations). 
They ranged from specific named entities (P4: \textit{``put it on \underline{the blue block}''}; P10: \textit{``move it to the \underline{yellow rectangle}''}) to category-level references (P1: \textit{``select all the \underline{earth blocks}''}; P4: \textit{``select all \underline{windmills}''}). \textbf{User as reference}, referencing the participant's own body, hand, or facing direction, appeared in 27.4\% of the referential operations. Common patterns included ego-relative directional terms (P5: \textit{``move it to \underline{the left}''}; P2: \textit{``move it \underline{backward from me}''}) and hand-as-reference expressions (P1: \textit{``place it at 0.5 meters in front of \underline{my hand}''}). \textbf{Environment as reference}, referencing absolute direction, position or scene structure, appeared in 14.8\% of the referential operations (P1: \textit{``align the bottom side with \underline{the ground}''}; P3: \textit{``place it on \underline{the floor}''}).

\paragraph{\textbf{\textit{Reference Frame}}}
Figure~~\ref{fig:findings_distribution}-c shows the identified Reference Frames as \textbf{object-centric, user-centric} and \textbf{environment-centric}. They appear to have a similar frequency distribution to Reference Source (object, user, environment). In addition, 10.2\% of the referential operations were coded as ambiguous.
We observed that when an Object served as the Source, the Frame was not always clear. 
While most object-sourced references adopted an object-centric frame (Levinson's \textit{intrinsic} frame), some could equally be interpreted through the user's viewpoint (Levinson's \textit{relative} frame). 
For example, \textit{``put it to the \underline{right} of the \underline{table}''} is ambiguous because ``right'' may refer to the table's canonical axis or the user's current facing direction. 

\paragraph{\textbf{\textit{Reference Anchor}}} 
We coded each referential operation for the specific attribute used to construct the spatial constraint (Figure~\ref{fig:findings_distribution}b).
\textbf{Geometric Anchors}---geometric properties such as surfaces, edges, centers, and orientation axes---were the most frequently observed (62.5\%), as expected from Tversky's finding that spatial descriptions preserve functionally relevant structure. Examples included surface placement (P3: \textit{``place it \underline{on top of} the red block''}), edge alignment (P8: \textit{``move it closer to \underline{the left edge}''}), and proximity (P5: \textit{``move the green block \underline{next to} the red one''}). \textbf{Semantic Anchors}---categorical or identity-based properties---appeared in 22.7\% of the referential operations, most prominently in Scenario~3 where labeled assets were available. Participants used color (P1: \textit{``select all \underline{the earth} blocks''}), type (P4: \textit{``select all \underline{windmills}''}), and identity labels (P8: \textit{``select the block with \underline{letter A}''}) to reference entities or groups.
\textbf{Historical Anchors}---past states of entities invoked as references for new operations---appeared in 14.8\% of the referential operations and emerged as a particularly noteworthy pattern. 10 out of 12 participants explicitly referenced prior states. These temporal references were fundamentally distinct from undo: when P10 said \textit{``return the three boxes to \underline{their initial positions}''} after multiple intervening operations, the intent was not to reverse a sequence but to selectively reference an earlier state as a target for a new transformation. This distinction---history-as-reference rather than history-as-reversal---was consistent across the operations involving a History Anchor.

\subsubsection{\textbf{\textit{Cross-reference strategies: Fluid frame switching and four compositional strategies (RQ2).}}}

Critically, participants fluidly switched among reference sources within the same manipulation sequence, often without explicit signaling, consistent with Levinson's documentation of implicit frame switching~\cite{levinson2003space}. For instance, P10 first used an Object Source (\textit{``move it onto \underline{the black box}''}), then a User Source (\textit{``move these three boxes closer to \underline{me}''}), and then an Environment Source (\textit{``drop it to \underline{ground}''}). Moreover, over 40\% of the referential operations involved combined references in which different source types were invoked within the same command---for example, P6: \textit{``line them up \underline{facing me}''} (User Source for orientation + Object Source for the group).
We observed a clear progression in referential complexity, from single-reference commands to multi-reference compositions 
(Figure~\ref{fig:findings_distribution}d). Our coding identified two dimensions of referencing strategies: \emph{Single} versus \emph{Multiple}, and \emph{Independent} versus \emph{History-dependent}, 
forming four strategies in combination.

\textbf{Independent Single Reference} uses one Source--Anchor pair to fully specify the intent, which was the most commonly seen 
(e.g., P11: \textit{``put this block \underline{on top of} the red block''}). \textbf{Iterative Single Reference} commands refined the target's own current state through repeated adjustments (e.g., P7: \textit{``make it bigger''} $\rightarrow$ \textit{``\underline{a bit} bigger''} $\rightarrow$ \textit{``stop''}); the implicit reference was the object's just-updated state, functioning as a Historical Anchor. 
\textbf{Parallel Multi-Reference} combined multiple reference sources within a single utterance, each placing a different constraint 
(e.g., P6: \textit{``line them up \underline{facing me}, in \underline{a row}''}---the row configuration referenced an Object Reference Source while \textit{``facing me''} referenced a User Reference Source). \textbf{Progressive Multi-Reference} sequences developed across multiple interaction turns, with each turn introducing a new reference that refined the cumulative result. A characteristic sequence from P8 illustrates this: \textit{``select this box''} $\rightarrow$ \textit{``move it to \underline{this position}''} [gesture] $\rightarrow$ \textit{``place it \underline{on top of that box}''} $\rightarrow$ \textit{``\underline{a bit} farther''} $\rightarrow$ \textit{``\underline{a bit more} to the left''} $\rightarrow$ \textit{``drop it''}---seven turns progressively accumulating spatial constraints from different references. 

Compositional complexity increased across scenarios: The proportion of system operations coded as Progressive Multi-Reference rose from 10\% in Task~1 to 37\% in Task~3, while those coded as Independent Single Reference decreased from 63\% to 28\%. 
Perhaps as participants became more comfortable with spatial manipulation with references, 
they exploited the full compositional capacity of referential language.

\subsubsection{\textbf{\textit{Resolving Implicity in referencial expressions using multi-modalities (RQ3).}}}

Consistent with Tversky's account of qualitative spatial language~\cite{tversky2005functional} and Oviatt's principle of complementary multimodal input~\cite{oviatt1999ten}, participants' commands ranged from fully explicit to partially implicit along different components (Figure~\ref{fig:findings_distribution}e). 
Our coding distinguished four types of Explicitness. 
 
Figure~\ref{fig:design_space} visualizes this space with examples. 
\textbf{Explicit} references fully specified Source, Anchor, and Frame (24.3\%; e.g., P11: \textit{``put my right-hand \underline{pointed} block at \underline{the center of} the scene''}). \textbf{Source Implicit} references (30.1\%) omitted the reference entity (e.g., P7: \textit{``make it bigger''}---bigger relative to what?), typically resolved through gesture or shared context. \textbf{Anchor Implicit} references (25.8\%) named the Source but left the magnitude qualitative (e.g., P3: \textit{``move it \underline{closer to} the front''}---how close?). \textbf{Frame Implicit} references (19.8\%) identified Source and Anchor but left the coordinate system ambiguous (e.g., P4: \textit{``move it to \underline{the right}''}---the user's right or the object's?).
 
Interestingly, participants tended to use different modalities or methods to resolve each type of implicity. Source-implicit ambiguity was most commonly resolved through gesture (pointing); Anchor-implicit commands tended to have iterative refinement (\textit{``\underline{a bit more},''} \textit{``too much, \underline{go back}''}); and Frame-implicit commands through body orientation (the system was expected to interpret \textit{``right''} from the user's current facing direction). Such complementary specification across modalities is consistent with 
the theories stating speech and gesture serve complementary roles in spatial reference construction~\cite{kita2003does, oviatt1999ten}.

%% file: 5-Design_Sapce.tex

\section{Reference Interpretation Framework}
\label{sec:design_space}

\begin{figure*}[t]
  \centering
  \includegraphics[width=\textwidth]{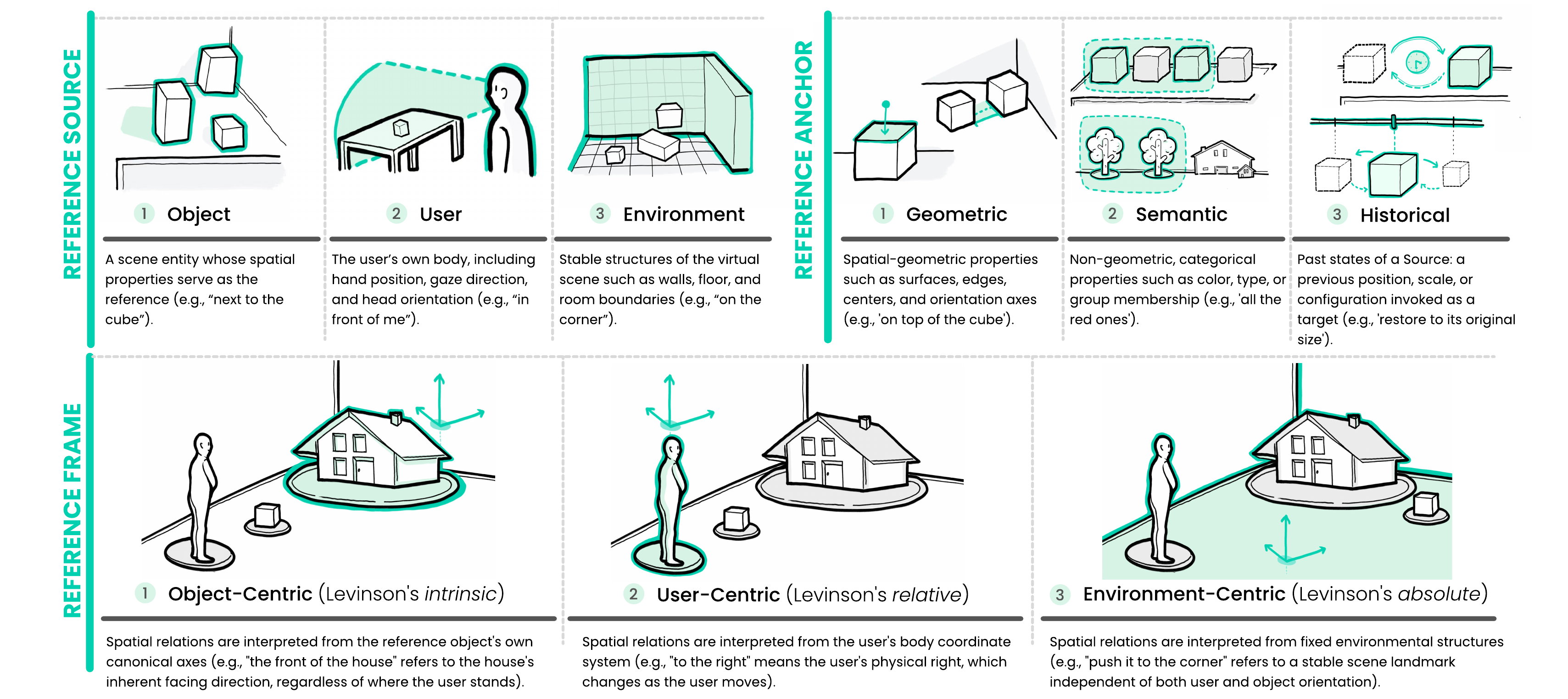}
  \caption{Reference Construction Model: $r$(Source, Anchor, Frame). The three components of a spatial reference: \textbf{Reference Source} (top left), \textbf{Reference Anchor} (top right), and \textbf{Reference Frame} (bottom). 
  }
  \Description{An illustrated reference showing three panels. The left panel depicts three Reference Source types (Object, User, Environment) with schematic illustrations. The center panel depicts three Reference Anchor types (Geometric, Semantic, Historical) with corresponding illustrations. The right panel depicts three Reference Frame types (Object-Centric, User-Centric, Environment-Centric) with illustrations showing how directional terms are interpreted differently under each frame.}
  \label{fig:reference_model}
\end{figure*}

The findings in the previous section establish that users' spatial intent in multimodal VR interaction is organized around references with identifiable internal structure (\textbf{Source, Anchor, Frame}) that vary in \textbf{Compositional Strategy} and information \textbf{Explicitness}.
This sets a foundation for us to build a conceptual framework for interpreting spatial references. This section illustrates this framework with 
a three-component reference construction model (Figure~\ref{fig:reference_model}). 

\subsection{Reference Construction Model} 
\label{sec:reference-model}

Each spatial reference can be decomposed into three components:

\begin{equation}
  \textbf{Reference} = r(\textbf{Source},\;\textbf{Anchor},\;\textbf{Frame})
  \label{eq:reference}
\end{equation}

\noindent \textbf{Source} is the entity providing spatial grounding; \textbf{Anchor} is the property extracted from that Source to construct a constraint; \textbf{Frame} is the cognitive coordinate system through which spatial relations are interpreted. Figure~\ref{fig:reference_model} defines the three categories within each component. 
While the three Reference Frames we defined use different terms for practical applicability, especially for system engineering, their definitions follow Levinson's ``Intrinsic, Relative and Absolute'' categorization. Extending beyond this, we define \textbf{Anchor} as an important dimension for RBM, and contribute its categorization (Geometric, Semantic and Historical).

\subsection{Compositional Strategies}
\label{sec:ds-matrix}

Our identified compositional strategies across references are another important part of reference interpretation. 
Figure~\ref{fig:design_space} illustrates a two-dimensional design space showing how references can be combined independently or progressively and how each component may be left implicit.
Progressive ambiguity arises when a new instruction may attach to different elements of prior context, such as the active target, an earlier reference, or an accumulated constraint. Accordingly, Independent–Explicit references require straightforward grounding, whereas Progressive–Implicit references require cross-turn state maintenance and contextual inference.
The next section introduces how our RBM pipeline decomposes inference into sequential Source, Anchor, and Frame recognition steps
.

%% file: 6-Pipeline.tex
\section{Reference-Based Manipulation}\label{sec:pipeline}

This section introduces our reference-based manipulation pipeline, \pipelinename{}, a pipeline for reference-based object manipulation in immersive environments. The system is designed to support object manipulation in ways that better align with human spatial cognition, enabling users to issue multimodal instructions based on references, such as surrounding objects, world information, user position and orientation, or prior object states, rather than relying only on absolute coordinates or fully explicit specifications. 

\subsection{\pipelinename{} Pipeline}

Our pipeline processes speech and gestural user input and scene context into executable manipulation actions through three sequential modules: 1) Perception, which captures user and scene data and organizes them into a queryable state representation; 2) Intent Recognition and Reference Construction, which identify the intended operation and target and construct the corresponding reference information; 
and 3) Execution, which retrieves reference-related data and translates them into executable parameters and constraints after confirmation. 
In this way, \pipelinename{} supports not only direct and explicit reference-based manipulation, but also more complex cases such as implicit referencing and progressive refinement across turns. 


\subsubsection{Perception Module}
The perception module translates raw sensor data into a structured JSON representation, categorizing the interaction context into three reference frames: user-centric, object-centric, and environment-centric. For user-centric data, the module tracks viewpoint orientation, hand positions, and ray-cast hits, allowing the system to interpret deictic gestures and egocentric references. Object-centric data includes current spatial attributes and geometric metadata, specifically the six bounding-box face centers, which serve as foundational anchors for manipulation. Environment-centric data defines fixed semantic anchors, such as cardinal directions and architectural boundaries. Furthermore, the module preserves interaction history and speech transcripts, providing the necessary temporal context to resolve implicit references across turns.

\subsubsection{Reference Analysis Module} 
After receiving the queryable data generated by the perception module, this module analyzes the user's multimodal input and constructs a reference with specified Source, Frame and Anchor. 


\paragraph{User Intent Recognition}
The step identifies the user's intended operation, or sequence of operations, together with target object(s) to which each operation applies. 
Based on the findings from our WoZ study, 
we defined five operation types: \emph{move, rotate, scale, color}, and \emph{copy} for reference recognition. \emph{Delete} was not included as the user command usually did not include references.
\paragraph{Reference Construction}
This step resolves the specific \textit{reference source}, \textit{frame}, and \textit{anchor} required to interpret the operation. A significant challenge in this process is the implicit information discussed in Section~\ref{sec:ds-matrix}, reference source and frame may be omitted and inferred from context, and anchors are frequently context-dependent. To address this, our pipeline utilizes a unified grounding process guided by two manually curated Example Guidelines: \textbf{Reference Guideline} and \textbf{Anchor Guideline} (Are shown in supplementary materials), they are constructed by manually drafting structured examples based on the interaction techniques identified in our design space. These guidelines map linguistic expressions to specific spatial logic, in reference guideline, each entry specifies a user instruction, its identified reference source, the associated frame (environment, user, or object-centric), and description of the reason of sources and frame. Anchor guideline change reference information to anchor information, and the description is about the reason of using this anchor.
To maintain LLM's output quality, our pipeline takes two steps to construct the reference for each user instruction: one for recognizing the Source and Frame, and the other for the Anchor. We separate these two steps because Anchor recognition was substantially less stable than Source/Frame recognition. Compared with frame recognition, which only needs to distinguish among three frame types, and source recognition, which can be supported by scene filtering and embodied cues, Anchor recognition must handle a wider variety of anchor types and often relies more heavily on speech instruction analysis.

This module supports both iterative and progressive refinement by maintaining two levels of records: object-level temporal states (position, rotation, etc.) and operation-level logs of previously executed instructions.
For iterative commands (e.g., ``a bit more to the right''), the system preserves the previously recognized reference state, allowing new inputs to be interpreted relative to the current frame. For progressive refinement, the module maintains target objects and prior constraints in a CSV file, enabling the system to interpret subsequent user instructions as additions to the existing spatial model. 
For each new instruction, the LLM jointly considers the current input and this structured history to infer whether it continues an active target or reference, refines an existing state, or adds a new constraint.
When a user instruction refers to a prior state, the system retrieves the relevant history record and treats it as an anchor for subsequent grounding and execution.

\begin{figure}
  \centering
  \includegraphics[width = 0.48\textwidth]{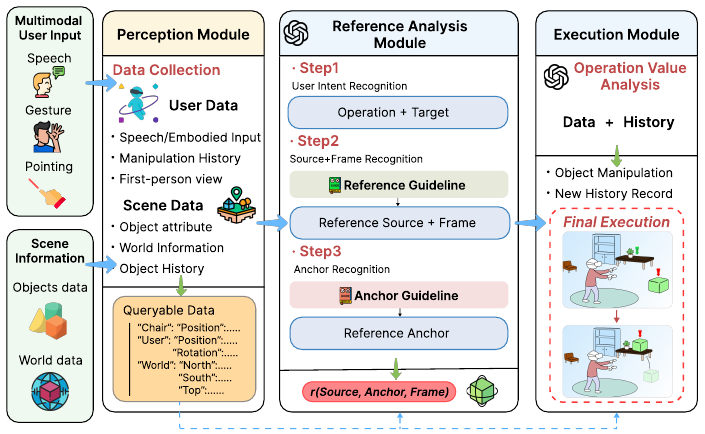}
  \caption{\pipelinename{} Pipeline: (1) Perception Module collects multimodal input and context into queryable data; (2) Reference Analysis Module identifies operations, reference sources, frames, and anchors; (3) Execution Module combines reference with queryable data and history to compute operation parameters and execute them. } 
  \Description{\pipelinename{} Architecture.}
  \label{fig:Pipeline}
\end{figure}

\begin{figure*}
  \centering
  \includegraphics[width=2\columnwidth]{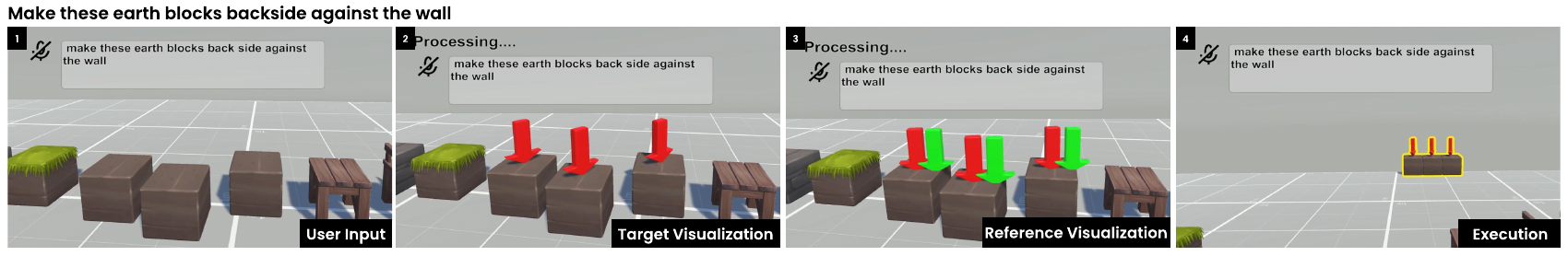}
  \caption{Example technique: ``Make these earth blocks' backside against the wall''. The system indicates interpreted target objects with animated red arrows, reference objects with green arrows, and highlights manipulated objects in yellow.}
  \label{fig:Earth_back_wall}
\end{figure*}

\subsubsection{Execution Module} 
After receiving the confirmed operations, targets, and references information, execution module extracts the anchor-related attributes required for execution from the queryable state space, such as position, size, bounding box, and orientation, etc. This data is then combined with the parsed operations to generate executable manipulation parameters and constraints. When history anchor 
is involved, the execution module additionally retrieves and includes the relevant records from the stored history file for computation. Once parameters are generated, they are applied to the target object(s) in the scene to complete the manipulation.

\subsection{Example Interaction Techniques} \label{InteractionTechniques}


We showcase several representative interaction techniques (listed in Section~\ref{appendix:InteractionTechniques}) enabled by \pipelinename{}. These techniques demonstrate \pipelinename{}'s capability to support multimodal reference-based manipulation. 

\subsubsection{System Capabilities}
Demonstrated manipulations include operations with all \textbf{three reference frames}: moving objects to a position relative to another object-Figure~\ref{fig:Progressive}-1st line (object-centric) or the user themself-Figure~\ref{fig:Sand_In_Front} (user-centric), scaling an object to half its own size-Figure~\ref{fig:Shrink_Windmill}, moving objects against the wall (Figure~\ref{fig:Earth_back_wall}) (environment-centric). 
Figure~\ref{fig:Earth_back_wall} also shows that ~\pipelinename{} supports manipulations involving \textbf{multiple reference sources, mixed reference frames}, and simultaneous \textbf{operations on multiple objects}. In this example, the user combines object-centric frame (``block's backside'') with environment-centric frame (``wall''). The instruction involves multiple reference sources-earth blocks and the wall, requires the system to coordinate these references so that the backside of each target object is oriented against the wall. At the same time, ~\pipelinename{} identifies multiple earth blocks as targets and performs multi-target operations.
Figure~\ref{fig:Progressive} demonstrates \textbf{progressive manipulation} across three turns. After placing the house on the grass block, RBM resolves ``it'' in ``Rotate it to face me'' to the active target, the house, preserves the placement relation, and adds a user-centric orientation constraint. In ``Push them to the corner,'' ``them'' expands the target set to the house and grass block, while the corner introduces an environment-centric constraint.



\subsubsection{User Interaction} In our prototype, users interact with the system through unconstrained natural language speech input, together with predefined finger gestures to control the microphone and select objects with raycasting, and start the AI process. 
For object indication, each hand is associated with a pointing ray. These rays can be used to indicate manipulation targets, reference sources and position.
Users can select a target object by pointing at it with the right-hand ray and pinching the right thumb and index finger together. 
After finishing giving instructions with speech or speech+pointing, the user triggers system action 
by pinching the left thumb and index finger together. 
Visual feedback is provided for selected or system-interpreted target objects (eg., Figure~\ref{fig:Earth_back_wall}-2), reference sources (eg., Figure~\ref{fig:Earth_back_wall}-3) and manipulated objects after execution (eg., Figure~\ref{fig:Earth_back_wall}-4). In addition, the system shows a circular animated glow on the ground when the reference frame is interpreted as the user themself (eg., Figure~\ref{fig:Sand_In_Front}-3).

\subsubsection{Implementation Details}
The demo system is developed on Meta Quest 3. The pipeline uses OpenAI API with GPT-4o (Large Vision-Language Model, LVLM) for analyzing operations, references, and attribute values. The system is deployed on a Windows 11 desktop equipped with an Intel Core i7-13650HX CPU, 32 GB of RAM, and an NVIDIA GeForce RTX 4060 GPU. 
The frontend is implemented in Unity 2022.3.27f1 with the SteamVR 2.8.0 SDK, which supports VR interaction, gesture recognition, and gesture-to-function binding. Speech input is triggered by a clenched-fist gesture, and speech-to-text processing are handled by Recognissimo\footnote{\url{https://bluezzzy.github.io/recognissimo-docs/demo/index.html}}, an offline speech recognition plugin for Unity. The system prompt are shown in our supplementary materials.

\subsection{Preliminary Technical Evaluation}
We conducted a preliminary technical evaluation to examine whether \pipelinename{} provides additional value beyond the reasoning capability of a general-purpose LLM. We compared the complete \pipelinename{} pipeline with a non-\pipelinename{} baseline, 
which is implemented by removing the reference-construction and reference-reasoning components from our pipeline. The baseline received the same user input and scene information as \pipelinename{}, then directly prompted the LLM to generate the manipulation operation, target object or objects, and corresponding execution parameters.
We evaluated both pipelines on two LLMs: GPT-5.5 as the most capable model available at the time of testing; GPT-4o as a lighter model for faster response.


Three instructions were tested:
\begin{itemize}
    \item [I1] ``Put the sand cube in front of me'', a single-target, explicit reference instruction.
    \item [I2] ``Make these earth blocks’ back side against the wall'', an instruction with multi-targets, multi- and implicit references.  
    \item [I3] A progressive, multi-reference instruction across multiple turns: 
    ``Place the bridge on the grass block;'' ``Place the house on the earth block;'' ``Shrink the house to half its size;'' ``Return it (the bridge) to its original position.'' 
\end{itemize}


Each model–pipeline–instruction combination was repeated ten times. This yielded 200 single-object manipulation outcomes in total. 
An outcome was counted as an error when the system selected the wrong operation, misidentified a target, resolved the wrong historical referent, or generated a result outside the predefined positional tolerance. End-to-end execution time was also recorded.

\subsubsection{Results}
\begin{table}[htbp]
\centering
\small
\vspace{-0.3cm}
\setlength{\tabcolsep}{3pt}
\caption{Error rates of RBM vs. Non-RBM across instructions using two LLMs. Average response time per execution.}
\label{tab:technical-evaluation}
\begin{tabular}{llccccc}
\toprule
Model & Pipeline & \textit{Err.}I1 & \textit{Err.}I2 & \textit{Err.}I3 & \textit{Err.}Total & Time (s) \\
\midrule
GPT-4o
& RBM
& 30.0\%
& 36.7\%
& 0.0\%
& 28.0\%
& 9.55 \\

GPT-4o
& Non-RBM
& 40.0\%
& 53.3\%
& 20.0\%
& 44.0\%
& 5.16 \\

GPT-5.5
& RBM
& 10.0\%
& 0.0\%
& 0.0\%
& 2.0\%
& 53.57 \\

GPT-5.5
& Non-RBM
& 10.0\%
& 30.0\%
& 0.0\%
& 20.0\%
& 23.89 \\
\bottomrule
\end{tabular}
\end{table}
\vspace{-0.2cm}
%

As shown in Table~\ref{tab:technical-evaluation}, \pipelinename{} reduced total error rates for both GPT-4o (from 44\% to 28\%) and GPT-5.5 models (from 20\% to 2\%). We can also see the largest benefit on I2 (30\% less error with GPT-5.5 and 16.6\% with GPT-4o) -- the implicit multi-object multi-reference instruction, 
where \pipelinename{} reduced errors in operation type, target identification, and spatial relation. RBM also improved error rates by 20\% for progressive instruction with GPT-4o but not with GPT-5.5. The difference was smaller for the simplest instruction I1 with both models. Therefore RMB is more valuable when instructions involve more complex referential strategies and / or when using less powerful LLMs for faster execution.

In our tests, the more powerful LLM achieved higher accuracy with/without RBM, but at cost of substantially longer execution time. RBM itself also increased execution time.
It's worth noting that the network conditions and cloud-server availability of online models may also have affected the measurements. Our implementation of \pipelinename{} techniques used GPT-4o to balance interpretation robustness and interaction responsiveness.

\section{Discussion}

From implementing ~\pipelinename{} as a working prototype, we learned lessons and observed several practical challenges. These experiences highlight which parts of reference-based multimodal manipulation are already tractable, and which remain difficult in our current implementation. 
We reflect on these experiences while suggesting potential solutions and future research. 

\subsection{Lessons Learned and Paths Forward}

\subsubsection{Reference recognition improved by example guidelines}
In early iterations, we found that reference and anchor recognition by LLMs was often unstable, especially for implicit or underspecified commands. We therefore introduced manually authored reference guidelines to provide structured examples for common reference patterns and anchor mappings. This visibly improved recognition stability. We make our example-based guidelines, manually curated from the data of the WoZ study, open-source as supplementary material. 
However, the system could not reliably interpret users' utterances when they are overly ambiguous or unstructured. 
For example, ``\textit{Drop it in the middle down there}'' without a complementary gesture could be interpreted in multiple ways. ``\textit{Uh, just put it, like, over there, ahead of me.}'' has little structure to be interpreted reliably. 
While guideline-based prompting improves stability for structured instructions, it remains sensitive to both structural variation and colloquial phrasing in open-ended user input. 
A potential solution is to add an instruction pre-processing step to the pipeline, which first transforms open speech 
into a more structured, command-like form before passing it to the reference analysis stages. In addition, as human utterances can be unavoidably ambiguous, supporting iterative intent refinement or simply making trial-and-error easy at the interface level is important.  

\subsubsection{Facilitating history retrieval}
Treating history as an anchor enables \pipelinename{} pipeline to support instructions such as ``Rotate according to its previous state,'' or ``restore it to its original size.'' 
In practice, however, the challenge is deciding which prior state should be treated as the intended anchor and how to identify the specific history. Depending on the instruction, the relevant history may be the original state, the immediately previous state, or an earlier state associated with a specific property such as size, rotation, position or even time event. We observed that users rarely remember exactly when each step took place, and their descriptions of past events are often vague. A possible direction is to investigate how users naturally describe historical states when using history as a reference anchor. 
Based on these patterns, future systems could assign semantic temporal labels to history records when each operation is stored, and update these labels dynamically as interaction progresses. For instance, the most recent action could be tagged as the previous step, while older records could be re-labeled according to their changing temporal relation to the current state. This could help narrow the history retrieval range and improve the precision of temporal anchor matching.

\subsubsection{Visualizing referential relationships}
Our current prototype visualizes the interpreted targets and reference sources with animated highlights. We find it important to convey such system status, especially as the pipeline takes a few seconds to execute. While highlighting objects or user as reference sources is straightforward, visualizing anchors, environment-centric references and relationships between multiple references requires more deliberate design solutions. 
For example, references such as ``the north,'' or ``world sky,'' may not correspond to a single entity in the scene. 
In multi-reference cases, the challenge is not only showing each reference independently, but also showing how they work together, such as when the moving direction is determined by the user's viewpoint and an object provides the start point. 
It also remains an open question how much underlying spatial reasoning should be shown to the user. While visual feedback of system status is important for human-AI alignment and error prevention / recovery, showing users a complex referential system can be distracting, as spatial reasoning happens to them rather intuitively. 

\subsubsection{Translating qualitative commands to parametric execution}

\pipelinename{} pipeline can reliably execute explicit scene manipulation by retrieving object positions, world directional points, and user state from the queryable representation. However, the system is less reliable when executing qualitative commands that require 
geometric reasoning, such as expressions with ``near,'' ``on top,'' ``at the edge,'' or more complex multi-reference constraints. These require contextual computation with the queried data and a good guess of scale based on the user and the task. 
Our suggestion is to separate LLM-based semantic interpretation from deterministic spatial computation. 
System engineers could design dedicated computational functions to handle such commands, rather than relying on a general-purpose model to infer both the meaning and the geometry at once. 
However, it can be hard to cover all the natural language requests with dedicated functions. For instance, higher-level instructions we got from the study such as ``Line them up facing me'' is a simple and intuitive command for the participant, but it requires complex reasoning of multiple reference sources and frames with additional constraints (alignment, distribution). Future work could introduce solutions to break down such requests into modular steps.

\subsection{Limitations}

Although \pipelinename{} benefits from a structured multi-stage pipeline, this design also introduces noticeable latency during interaction. In our current prototype, 
each AI-supported stage typically takes about 2.5-5 seconds,
leading to an end-to-end process 
typically taking 10-20 seconds. 
Major bottlenecks include internet communication cost of API and the GPT model’s processing of visual input, which is generally slower than text-only reasoning. Several directions could reduce this latency. First, locally deployed lightweight models could reduce communication overhead and be fine-tuned specifically for reference-based manipulation, providing a better balance between speed and accuracy. Second, some visual processing could be moved outside the LLM loop: task-specific semantic segmentation and spatial-relation models could convert images into structured object metadata or a scene graph in advance, allowing LLM to reason primarily over text and structured spatial information rather than repeatedly analyzing raw images. At the interaction level, users could continue other operations while the system processes a request and receive a visual notification when the result is ready. These approaches could reduce both actual and perceived waiting time in future implementations.

Our current handling of progressive ambiguity is limited by its flat structured history. Although it supports short sequences like in Figure 10, it has not been systematically tested for longer open-ended tasks, where many prior targets, references, or constraints may remain plausible. 
The current pipeline was designed to examine the framework’s feasibility rather than to support naturalistic task use. Doing so will require hierarchical context management and tracking of active targets and constraint dependencies.

We also acknowledge that our manually curated example guidelines are experimental, serving as a starting point. Future iterations could consider learning these mappings from data and employing more robust few-shot prompting techniques. 
Additionally, our WoZ methodology enabled unconstrained observation but also introduced the wizard's interpretive biases: response delays and occasional misinterpretations may have affected participants' strategies. Human wizard's intelligent capacity to interpret vague and high-level instructions 
may have inflated the referential complexity. 
Future work will test our working system to evaluate the user experience and observe users' adaptive behavior. 
Our study tasks were also constrained to a fixed set of basic operations; manipulation types not covered, such as deformation, physics-based interaction, or collaborative multi-user coordination, may elicit referential patterns the current framework does not capture and call for more diverse gestural integration. Extending the task repertoire would enrich the conceptual framework.  

%% file: 9-Conclusion.tex
\section{Conclusion}

This paper introduces Reference-Based Manipulation (RBM), a framework and pipeline that reorients multimodal interaction for object manipulation in immersive platforms around the construct of spatial reference. 
Grounded in spatial cognition theory and a WoZ study of unconstrained user behavior, we decomposed spatial intent into three core components—Source, Anchor, and Frame—and characterized how users combine references through different strategies and degrees of explicitness. We demonstrated the viability of this approach through an LLM-based pipeline that processes speech and gesture input to construct and execute spatial references. Our implementation supports a range of interaction techniques, including multi-source, multi-frame, and temporally progressive manipulations. By treating reference construction as the central design challenge, RBM offers a pathway toward more natural, flexible, and cognitively aligned interfaces for immersive 3D environments. We hope this work provides both a conceptual foundation and a practical starting point for future systems that aim to understand and act upon how humans naturally communicate spatial intent.

%% file: 10-Appendix.tex
\clearpage
\onecolumn
\appendix
\section{Appendix}


\subsection{Overview and comparison of related techniques}
\label{appendix:comparison}

\begin{table}[!htbp]
\caption{Comparison of recent speech-based multimodal interfaces for spatial interaction in XR.}
\label{tab:comparison}
\centering
\small
\renewcommand{\arraystretch}{1.35}
\setlength{\tabcolsep}{8pt}

\begin{tabular}{
@{}
p{0.23\textwidth}
p{0.34\textwidth}
p{0.35\textwidth}
@{}
}
\toprule
\textbf{System} &
\textbf{Semantic Analysis (LLM)} &
\textbf{Spatial Reference} \\
\midrule

\raggedright VRCopilot~\cite{zhang2024vrcopilot} &
\raggedright Intent parsing &
\raggedright --- 
\tabularnewline

\raggedright GazePointAR~\cite{lee2024gazepointar} &
\raggedright Pronoun disambiguation &
\raggedright Pronoun resolution
\tabularnewline

\raggedright Ostaad~\cite{aghel2024people} &
\raggedright Observational &
\raggedright Observed, not modeled
\tabularnewline

\raggedright GesPrompt~\cite{hu2025gesprompt} &
\raggedright Parameter extraction &
\raggedright ---
\tabularnewline

\raggedright EchoLadder~\cite{hou2025echoladder} &
\raggedright Suggestion generation &
\raggedright ---
\tabularnewline

\raggedright VR Mover~\cite{wang2025can} &
\raggedright Intent parsing &
\raggedright Context tracking
\tabularnewline

\raggedright ASSISTVR~\cite{chen2025analyzing} &
\raggedright Intent classification &
\raggedright ---
\tabularnewline

\raggedright \textbf{RBM (Ours)} &
\raggedright \textbf{Reference reasoning} &
\raggedright \textbf{\textit{r}(Source, Anchor, Frame)}
\tabularnewline

\bottomrule
\end{tabular}
\end{table}

\subsection{Examples for referencing strategies and implicit commands }

\begin{figure*}[htbp]
  \centering
  \includegraphics[width=\textwidth]{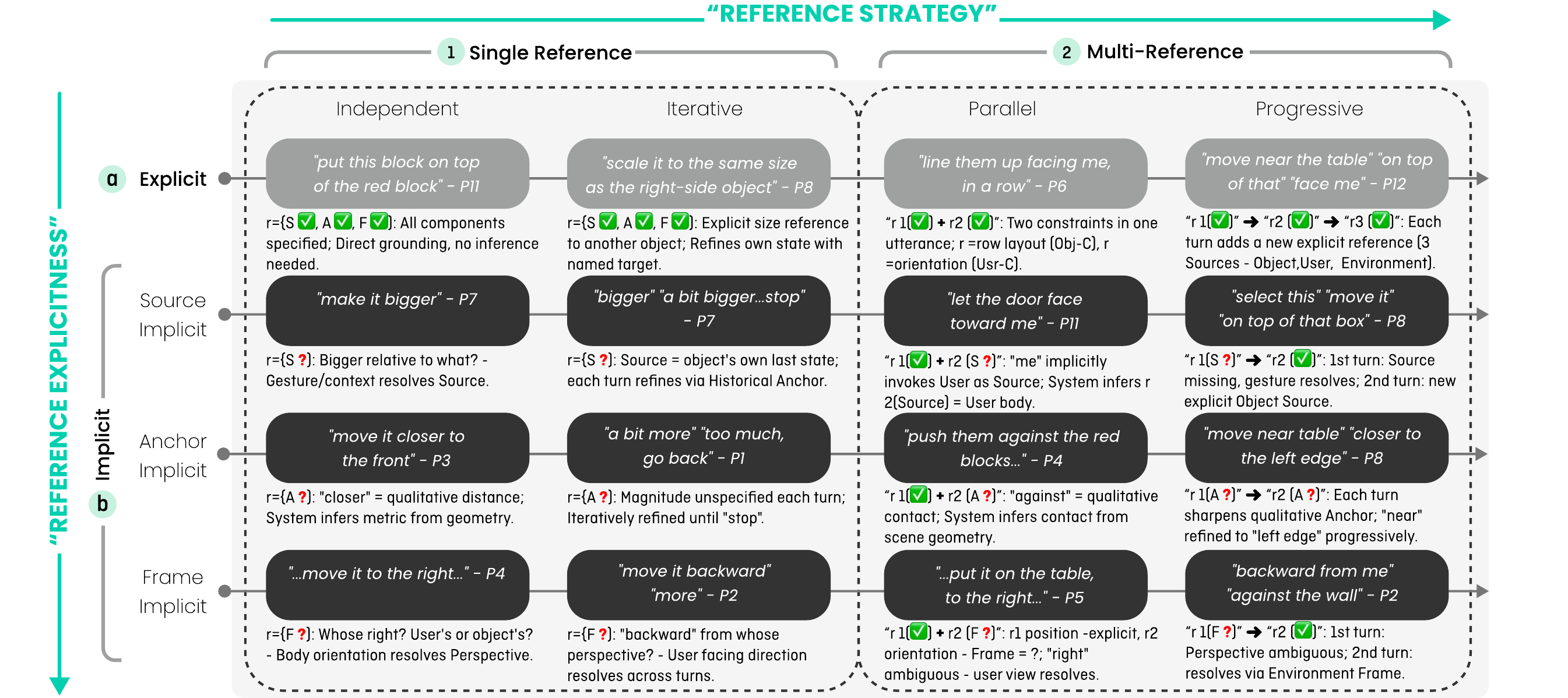}
  \caption{Referencing \textbf{Strategy} $\times$ \textbf{Explicitness} matrix. 
  \textsuperscript{*}One cell uses a synthesized example consistent with observed patterns.}
  \Description{A four-by-four matrix figure. Columns are labeled Independent, Iterative, Parallel, and Progressive. Rows are labeled Explicit, Source Implicit, Anchor Implicit, and Frame Implicit. Each of the sixteen cells contains a participant quote in italics, a reference decomposition notation showing which components are specified or missing, and a brief resolution description in teal text.}
  \label{fig:design_space}
\end{figure*}

\subsection{More Example Interaction Techniques}
\label{appendix:InteractionTechniques}

\begin{figure*}[htbp]
  \centering
  \includegraphics[width=\columnwidth]{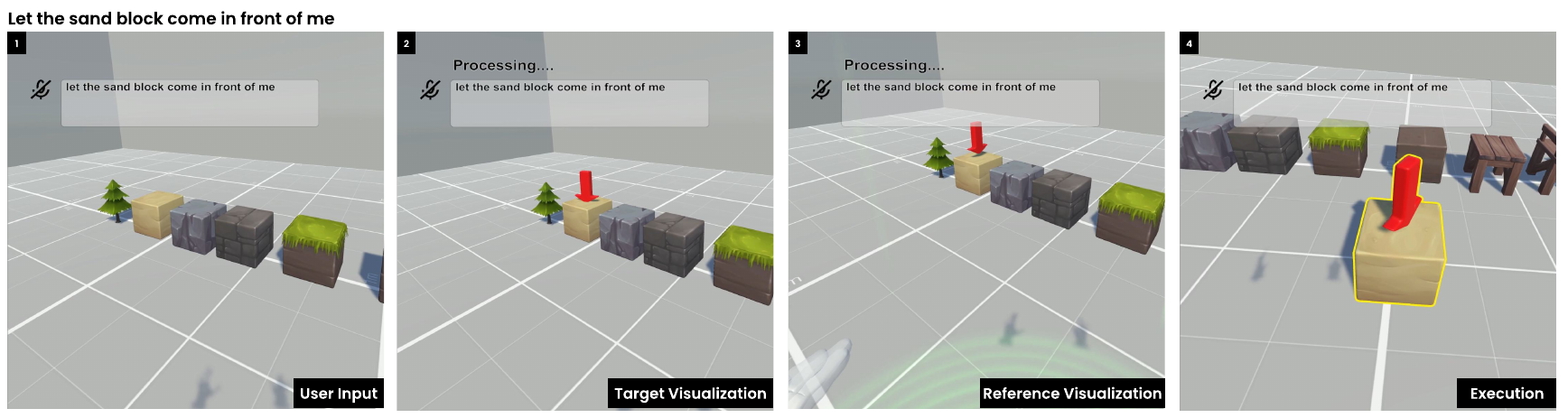}
  \caption{System interpreting and responding to user instruction:``Let the sand block come in front of me.'' Red arrow highlights the target; Green light ring indicates the user as reference source.}
  \label{fig:Sand_In_Front}
\end{figure*}

\begin{figure*}[htbp]
  \centering
  \includegraphics[width=\columnwidth]{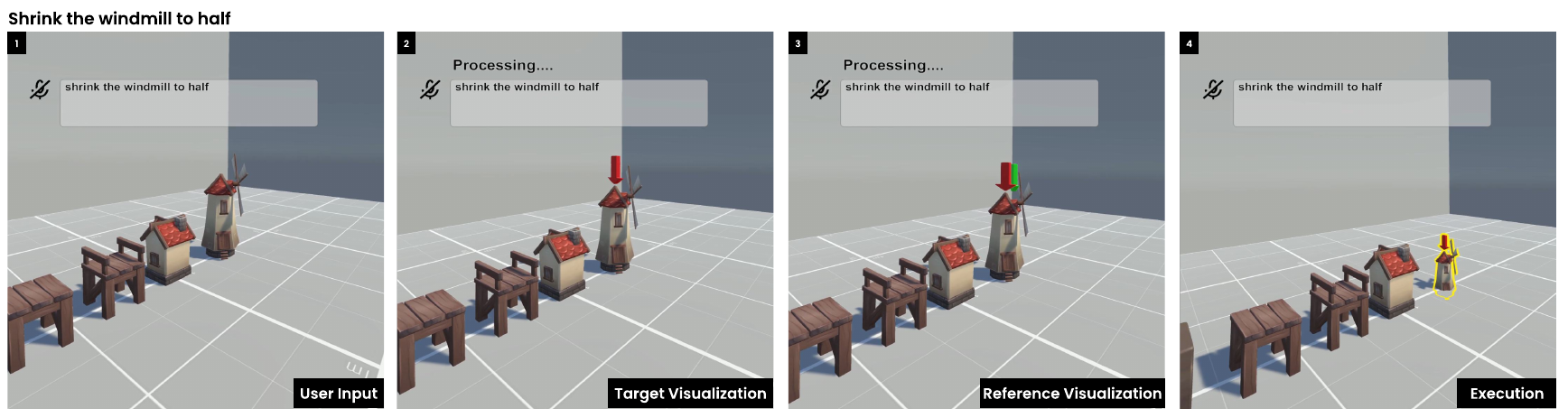}
  \caption{User instruction: ``Shrink the windmill to half.'' The windmill is interpreted both as the target and the reference source.}
  \label{fig:Shrink_Windmill}
\end{figure*}


\begin{figure*}[htbp]
  \centering
  \includegraphics[width=0.985\columnwidth]{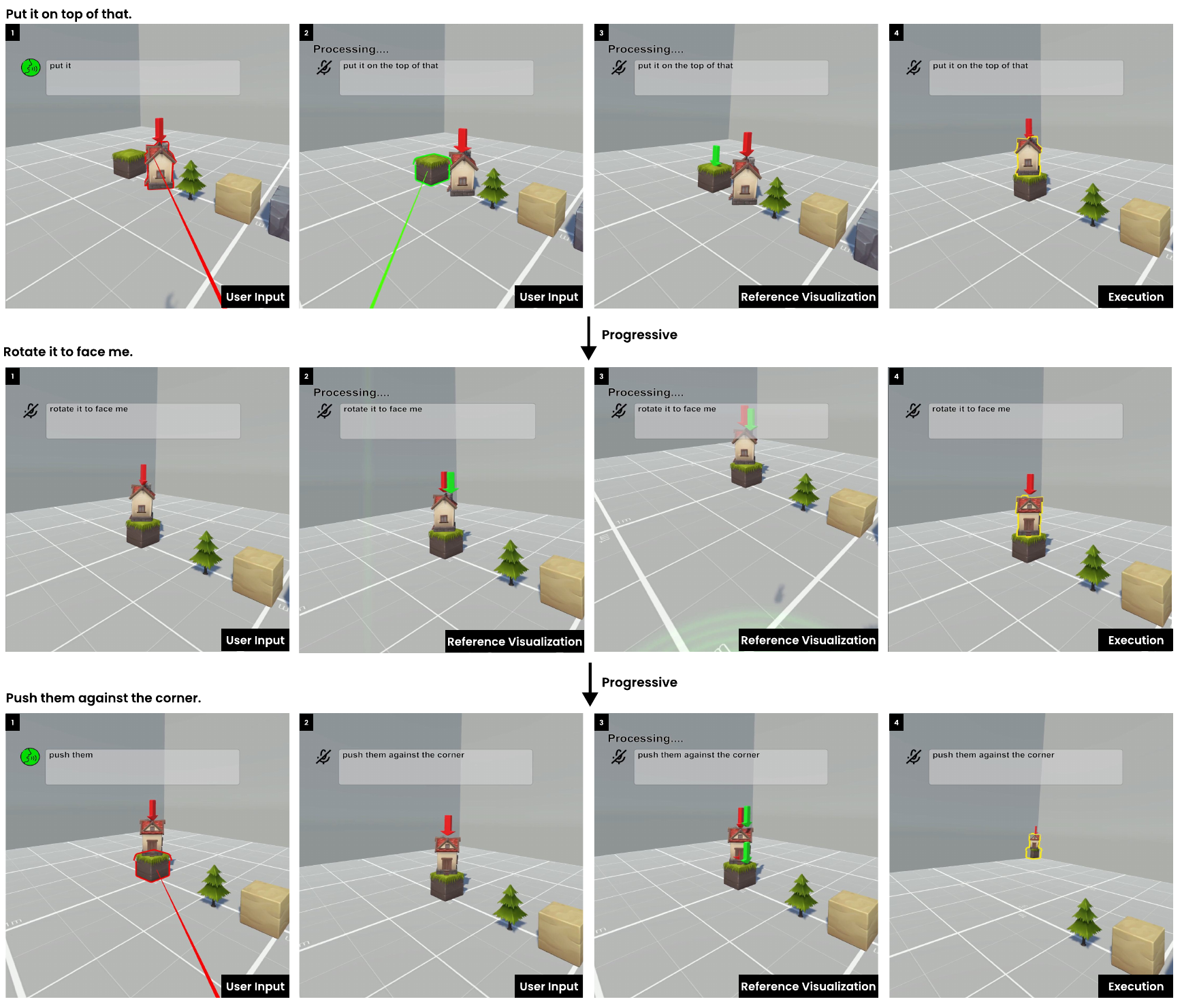}
  \caption{Progressive Manipulation: (Instruction 1) ``Put it (red ray pointing) on top of that (green ray pointing)''; (Instruction 2) ``Rotate it (refer to the previous target) to face me''; (Instruction 3) ``Push them (red ray pointing) against the corner''.}
  \label{fig:Progressive}
\end{figure*}